\documentclass[twocolumn,showpacs,preprintnumbers,amsmath,amssymb,showkeys]{revtex4}
\usepackage{graphicx}
\usepackage{dcolumn}
\usepackage{bm}
\begin{document}
\title{Ferromagnetism in one dimension: Critical Temperature}
 \author{S. Curilef}
\author{L.A. del Pino}
\altaffiliation[Permanent Address: ]{Facultad de Monta\~na,
Universidad de Pinar del Rio, Cuba.}
\author{P. Orellana}
\affiliation{%
Departamento de F\'\i sica, Universidad Cat\'olica del Norte, Av. Angamos 0610, Antofagasta, Chile.}%
\date{\today}
\begin{abstract}
Ferromagnetism in one dimension is a novel observation which has
been reported in a recent work (P. Gambardella et.al., Nature {\bf
416}, 301 (2002)), and it is thought that anisotropy barriers are
responsible in that relevant effect. In the present work,
transitions between two different magnetic ordering phases are obtained as a result of an alternative approach.
The critical temperature has been estimated by
Binder method. Ferromagnetic long range interactions have been
included in a special Hamiltonian through a power law that decays at
large interparticle distance $r$ as $r^{-\alpha}$ for
$\alpha\geq0$.
We found that if the range of
interactions decreases ($\alpha\rightarrow\infty$), the trend of the
critical temperature disappears,  but if the range of interactions
increases ($\alpha\rightarrow0$),
the trend of the critical temperature approaches to the mean field approximation.
The crossover between two these limiting situations is discussed.
\end{abstract}
\maketitle

Recently, much attention has been paid to structures of lower
dimensionality\cite{prlDorantes,prbShen,PGaNat416,prlMermin,prbCannas,prb58,PGaprb61}.
As the space dimension of a physical system decreases, magnetic
ordering tends to vanish as fluctuations become relatively more
important. In particular, there are no spontaneous magnetization in
several one dimensional models, at any nonzero temperature; for
instance, the isotropic spin-S Heisenberg model with finite range
exchange interaction~\cite{prlMermin} and the classical gas model
with hard-core and finite range interactions\cite{Hove}. However,
anomalies such as anisotropy properties as microscopic long range
interactions are not taken into account at finite temperature.

 Regardless, ferromagnetism in one
dimension has been recently reported for monoatomic metal chain of
Co constructed on a Pt substrate, with anisotropy
barriers~\cite{PGaNat416}. It was found experimental evidence that
the monoatomic chains consist of thermally fluctuating segments of
ferromagnetically coupled atoms. Chains evolve into a ferromagnetic
long range ordered state owing to the presence of anisotropy
barriers below a threshold temperature~\cite{PGaNat416}.

Much effort has been devoted to handling finite and infinite range
interactions in computational systems by molecular dynamics and
Monte Carlo simulations due to the absence of exact and analytical
results. However, some special situations in one dimension can be
studied exactly; for instance, \begin{itemize} \item mean field
theory (this is, if the range of interactions was infinite) and
coupling to first neighbors\cite{Reichl},
\item $1/r$ and logarithmic potentials in a periodic media by infinite
repetition of a central cell\cite{SCuExact}, \item hard core
classical gas\cite{Hove}, \item isotropic spin-s
model~\cite{prlMermin},
\item the thermodynamics of the Casimir force and the excess free
energy of d-dimensional spherical model \cite{pre70} and \item a
finite size scaling theory is developed when a particular family of
interactions decays slowly with the distance \cite{mplb17}, etc.
\end{itemize}

In this paper, we present a novel approach to ferromagnetism in one
dimension. More explicitly, the presence of infinite range
microscopic interactions induces some important modifications to the
thermodynamic properties of systems, some of them have still been
not characterized. Several evidences of a ferromagnetic state have
been suggested due to the existence of long range interactions in
some physical systems in one
dimension~\cite{prb56,prl86,prbCannas,prb58,prl76,prb34,pre70,mplb17,pra45}.

The main aim of the present work is to obtain the critical
temperature between two states of different magnetic ordering in a
spin-$\frac{1}{2}$ Ising linear chain, where the range of
interactions is, at least, comparable to the size of the chain. It
is well known that there is a state of magnetic ordering in the mean
field approximation. That approximation focusses on a single
particle and assumes that the most important contribution to the
interaction of such particle with all particles is determined by the
mean field due to other particles. The configuration with lowest
energy is one in which the spins are totally aligned. Before, it was
explained that no magnetic ordering is observed for finite range of
interaction ({\em e.g.,} first neighbors); but, which happens
between infinite and finite range of interactions will be
illustrated through a generic power law decaying as $1/r^\alpha$,
where $r$ is the distance between two particles and $0\leq \alpha <
\infty$, ($\alpha=0$ and $\alpha \rightarrow \infty$ close
respectively to mean field and to independent spin approximations).
Hence, the principal question that we try to answer in the present
work is ``How does the critical temperature depend on the range of
the interaction?"

Firstly, the critical temperature for a particular case (namely,
$\alpha=2$) of this kind of systems has been earlier reported by
various authors and by the use of several
techniques\cite{JPC3,JPC4}. A comparison of results is done in a
previous work, see for instance Ref.\cite{cssXII}. Secondly, the
critical temperature as a function of other values of $\alpha$
(namely, $1<\alpha\leq 3/2$) was reported in Ref.\cite{SCaIJMPC}.
The critical temperature $T_c$ tends to infinite as $\alpha $ tends
to $1^+$ (this is, $\lim_{\alpha\rightarrow 1} T_c \rightarrow
\infty$) in the thermodynamic limit. In general, a strong dependence on
the size of the system has been observed for such model. Thirdly,
the thermodynamic limit is not reached for $\alpha \equiv 1$.

So that, no results have been previously reported for $\alpha \leq 1$.
 In the present work, we propose a model for this kind of
 system which allows to calculate the critical temperature by
 standard methods in the thermodynamic limit for $0\leq \alpha <\infty$. We carry out
a numerical study through Monte Carlo procedure and we report direct
results from our model which considers information on the range of
interactions.

The system is described by the following Hamiltonian
\begin{equation}
H = -  \sum_{i,j =1}^n J(i-j) s_is_j \label{Hs}
\end{equation}
where   $s_i= \pm 1 \;\;\;\forall i$ and $(i-j)$ is the distance
between two sites and $n$ is the number of particles in every cell.
\begin{equation}
J(K)=\frac{1}{2}\sum_{l=-L}^{l=L}
\frac{J_\alpha^L(n)}{|nl+K|^{\alpha}}
\end{equation}
where
\begin{equation}(2L+1)n=N \label{N}
\end{equation} is the total number of particles in $2L+1$ repetition of a
central configuration,
\begin{equation}
J_\alpha^L(n) =
\frac{J}{2^\alpha}\frac{1-\alpha}{N^{1-\alpha}-1}=\left\{
\begin{array}{ll} \frac{1-\alpha}{
2^\alpha}J / {N^{1-\alpha}} &\text{if $\alpha < 1$} \\ & \\
\frac{1}{ 2}J / \ln N &\text{if $\alpha = 1$}
\end {array} \right. \label{Hs4}
\end{equation}
where $J$ is a positive parameter, $J_\alpha^L(n)$ measures the
strength of the coupling that depends on the size $n$ of the system,
and $J_\alpha^L(n)\rightarrow(\alpha-1)J/2^\alpha$ if $\alpha
> 1$ \cite{note}.

Let us consider a computational cell in one dimension with size $n$.
Periodic boundary conditions have been applied through infinite
replications of a central cell. Recently, the problem related to the
periodic boundary conditions in systems with microscopic long-range
interactions has attracted the attention of several authors (see,
for instance\cite{Lekner,Jensen,efficient,SCaIJMPC,cssXII}).
Periodic boundary conditions were applied in a similar manner which
was recently discussed\cite{SCaIJMPC}. However, this way was already
introduced by Curilef\cite{SCuIJMPC}.

Thermodynamics describes the behavior of systems with many degrees
of freedom after they have reached a state of thermal equilibrium.
Furthermore, their thermodynamic state can be specified in terms of
a few parameters called state variables. At equilibrium, macroscopic
observables are linear functions of $n$ (number of particles). If
the function $f(n)$ is an observable, the thermodynamic properties
impose to observables to be a homogeneous linear function of $n$,
this is $f(n) = nf_n $, where $f_n = f(n)/n$ and $f(\lambda n) =
\lambda f(n)$ for very large $n$\cite{Reichl,Ruelle}. At this point,
it is important to remark that by choosing the Hamiltonian of
Eq.(\ref{Hs}) (that contains ferromagnetic interactions that decay
as a $1/r^\alpha$ law) we can find two facts: \begin{enumerate}
\item The nice property, about observables as a linear homogeneous
function of $n$, is verified for thermodynamic functions, e.g., the
internal energy.
\item The parameter $\lambda$ can be explicitly written from
Eq.(\ref{N}).\end{enumerate} As a consequence of the previous
properties, a weak dependence of the size of the system is expected
for results that come from this model.

\begin{figure} \centering
\includegraphics[angle=270,width=0.5\textwidth]{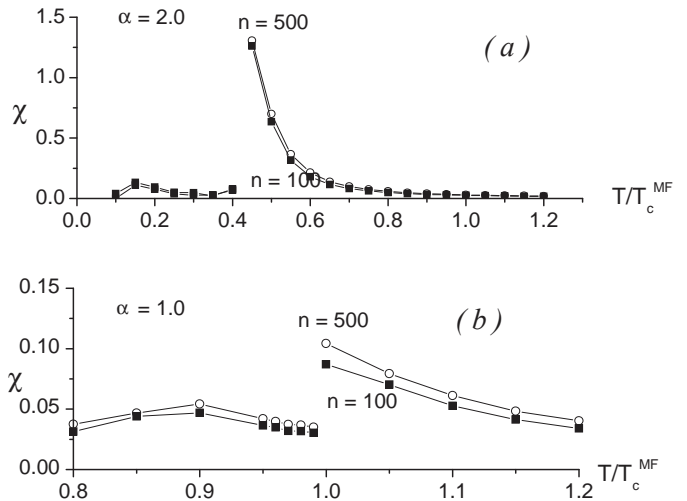}
\caption{\label{fig1} The susceptibility as a function of the
temperature $T$ is depicted by (a) $\alpha =2$ and (b) $\alpha =1$,
the values of $n$ are indicated in the figure.}
\end{figure}

The present computational study was carried out on a linear chain
with $L = 10^{6}$ and $n$ between $10^2$ and $10^3$, to give
effective sizes of the chain of the order from $10^8$ to $10^9$
particles in according to the Eq.(\ref{N}), and several values of
$\alpha$. Standard techniques are taken into account to compute its
normalized autocorrelation function $C(t)$, $C(t)\sim 1 \%$, due to
the large fluctuations near the critical point.

The magnetic susceptibility as a function of temperature is
estimated from the magnetization fluctuation,
\begin{equation}
\chi(T)  =\left\{
\begin{array}{ll} \frac{N}{k_B T}(\langle s^2 \rangle_n - \langle
|s|\rangle^2_n) &\text{for $T<T_c$}\\ &\\ \frac{N}{k_B T}\langle s^2
\rangle_n &\text{for $T>T_c$},
\end {array} \right.
\end{equation}
where $T_c$ is the critical temperature, $<s>=0$ for $T>T_c$.  In
the Fig.\ref{fig1} typical sets of curves for the magnetic
susceptibility are depicted as a function of a reduced temperature
$T/T_c^{MF}$, where $T_c^{MF}=J_\alpha^L(n)/k_B$, where $k_B$ is the
Boltzmann constant. Range of interactions is given by (a) $\alpha=2$
and (b) $\alpha=1$, and sizes of the central cell are shown for
$n=100$ and $500$, for $L=10^6$. The total number of particles is
given by Eq.(\ref{N}). The trend of the magnetic susceptibility
$\chi(T)$ suggests a discontinuity in Fig.\ref{fig1} and a change of
the magnetic ordering phase is related to such property. Circles are
obtained from simulations and lines correspond to a linear
interpolation of the data in the Fig.\ref{fig1}. It is expected that
the trend of the magnetic susceptibility becomes a behavior similar
to the mean field approximation, this is the magnetic susceptibility
has an infinite jump at the critical temperature.

\begin{figure} \centering
\includegraphics[angle=270,width=0.45\textwidth]{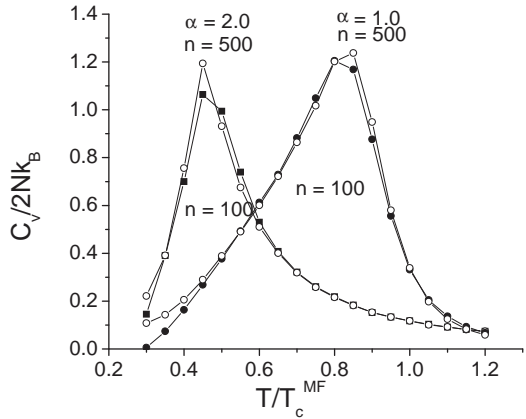}
\caption{\label{fig2} The specific heat as a function of the
temperature $T$ is shown. The trend of these functions shows also a
discontinuity. The values of $\alpha = 1$, $2$ and $n=100$, $500$
are indicated in the figure }
\end{figure}

The specific heat is estimated as a function of the temperature from
the energy fluctuation, which is given by
\begin{equation} C(T) = \frac{N}{T^2}\left(\langle H^2
\rangle_n - \langle H\rangle^2_n\right).
\end{equation} In Fig.\ref{fig2}, the typical trend of specific heat is depicted. Numerical calculations were carried out for several values of
$n$ and $\alpha$; namely, $n=100$, $500$ and $\alpha=1$, $2$.
A discontinuity is also observed for the specific heat at critical
temperature.

The phase transition can be characterized by several ways. A
suitable approach to define the critical point in finite system is
the Binder method. A typical property is looked for from the profile
of the fourth-order cumulant of the magnetization\cite{Binder}. A
standard behavior of the phase diagram is expected for this kind of
system. The Binder cumulant of fourth order is defined as
\begin{equation}
U_n=1-\frac{\langle s^4 \rangle_n }{3 \langle s^2 \rangle_n^{2}}.
\end{equation}

Cumulants $U_n$ as a function of the temperature, for several sizes
of system $n$, are intersected in a common point. Such point is the
critical temperature which depends on values of the parameter
$\alpha$. The typical behavior of the Binder cumulant $U_n$ (or
equivalent quantity $g_n=-3U_n$ called the renormalized coupling
constant\cite{prb32}) is rather stable under critical temperature;
however, fluctuations can be important above the critical point. So
that, this properties permits to distinguish the critical point by a
very simple graphical criteria.

\begin{figure} \centering
\includegraphics[angle=0,width=0.45\textwidth]{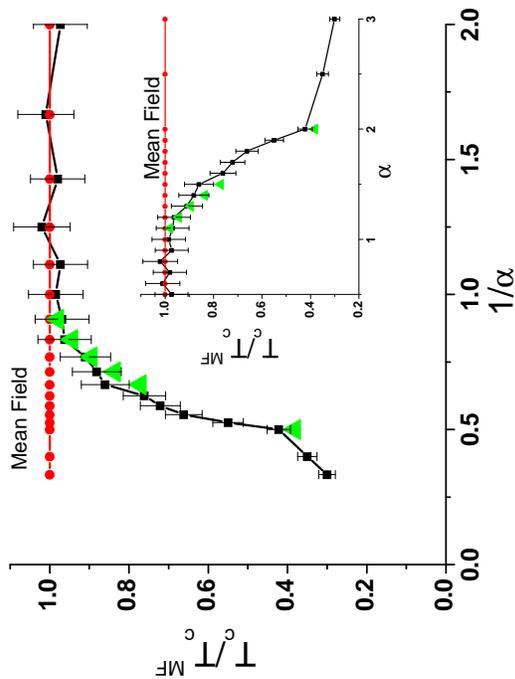}
\caption{\label{fig3} The critical temperature as a function of
$1/\alpha$, the inverse of the range of the microscopic interaction,
is depicted. The trend of $T_c$ goes to zero when $1/\alpha
\rightarrow 0$, namely $\alpha \rightarrow \infty$. The inset shows
the trend of $T_c$ as a function of $\alpha$. Error bars represent a
deviation of the 5 percent or less at each point of results.}
\end{figure}

In the Fig.\ref{fig3}, the critical temperature is depicted as a
function of $1/\alpha$ for the magnetic ordering transition in the
thermodynamic limit. Error bars are included to represent a 5
percent or less of the deviation for the obtained values at each
point of the critical temperature. Additionally, an inset is
included in Fig.\ref{fig3} to show the same fact for $0 \leq \alpha
\leq 3$. Both pictures are included to emphasize the following
features about critical temperature:
\begin{itemize}
 \item It is close to the mean field one for $\alpha
\rightarrow 0$.
\item It is lesser than the critical temperature given by the mean field approximation for $\alpha > 1$.
 \item It falls to zero for the short range interaction regime (e.g., nearest
neighbor) as it is expected.
\item It is depicted as a function of $1/\alpha\rightarrow 0$ to remark that goes to
zero while $\alpha\rightarrow \infty$.
\item It is depicted the crossover between two known limiting
cases.
\item It is recovered by the present model the known results for
particular values of $\alpha$; namely, $\alpha=1.1$, $1.2$ $1.3$,
$1.4$ and $1.5$\cite{cssXII} and $\alpha=2$\cite{SCaIJMPC}.
\end{itemize}

\begin{table}
\caption{Comparison of our estimation of the critical temperature to
previous result in the literature\cite{prb56}.}
\begin{tabular}{c |c| c}
  \hline
  $\alpha$ &$T_c$ This work&  $T_c=(1-\alpha)/2^\alpha K_c$  \\
  \hline \hline
  1.1 & 0.9689 & 0.9797309  \\
  1.2 & 0.9625 &0.9438766   \\
  1.3 & 0.9101   & 0.8951426   \\
  1.4 & 0.8813  & 0.8367191  \\
  1.5 & 0.86  & 0.7714285   \\
  \hline
\end{tabular}
\label{tabla}
\end{table}

We can compare the critical temperature with earlier results which
are presented in the literature. We have chosen the Ref.\cite{prb56}
because the authors have made an exhaustive search of the critical
couplings $K_c$ as a function of $\alpha$ and compare their values
to others\cite{tableIII}. In order to compare critical temperatures
it is necessary to make a simple transformation from Eq.(\ref{Hs4})
for $N\rightarrow\infty$ given by $T_c=(1-\alpha)/2^\alpha K_c$. The
comparison is in the Table~\ref{tabla}. In a similar way, for
$\alpha = 2$, our critical temperature is $0.4221$ and we can
compare it to a previous value $0.3816$ obtained from the
Ref.\cite{cssXII}.

In general, according to the Table~\ref{tabla}, our estimation for
the critical temperature is greater than the value obtained by other
author previously. As it is expected that in the limit of short
range interactions  $T_c \rightarrow 0$, it is reasonable to expect
that the lower value could be more acceptable. Simulations for
larger systems had been carried out in order to obtain accurate
results in one dimension\cite{prb56}. However, the numerical work is
much more costly due to the enormous number of particles ($n=300
000$) in systems, in contrast to the method which has been
implemented here, few particles in a cell and a big number of
repetitions over all space. If replicas of the central cell are on
increase, the time of computation does not practically suffer
changes, because such time only depends on the number of particles
on the central cell. Sometimes, computation facilities are not
sufficient for filling the demands of resources that make
simulations of many-body systems; then, it is very important to
resort to an alternative numerical way.

On the one hand, a simple theoretical argument given by L.
Reichl\cite{Reichl} shows that a finite Ising chain in one dimension
with a number of ferromagnetically coupled spins cannot exhibit a
phase transition. If the external magnetic field goes to zero the
order parameter $\langle s \rangle$ tends also to zero. Hence, no
spontaneous nonzero value of the order parameter is possible. On the
other hand, another simple theoretical argument given by the same
author\cite{Reichl} shows that the mean field theory predicts a
phase transition at a finite temperature for a lattice in one
dimension. Both arguments are not opposite between them. In the
present point of view, the mean field theory (e.g. $\alpha=0$)
approaches to the case of every spin interacts with each other
without discriminating over the sites and an Ising chain is the
limit whose interactions are very short ranged (e.g.
$\alpha=\infty$). In this work, we have discussed the crossover
between both limiting cases, with a system of finite
ferromagnetically coupled spins that obeys to the Hamiltonian given
by Eq.(\ref{Hs}).

Summarizing, the inhomogeneity which generally characterizes to
systems with interactions with $\alpha \leq 1$ is removed by the
scaling $J/J_\alpha^L(n)$ suggested in Eq.(\ref{Hs4}), which is used
to define the Eq.(\ref{Hs}). Thus, if we suppose that the  scaling
represents the number of nearest neighbor spins, the size of the
system (the total number of spins in the chain) is always greater
than the such scaling for $0 < \alpha \le 1$. Both values, nearest
neighbor and size of the system,  are coincident for $\alpha=0$. In
this way, we expected that a proper thermodynamic limit can be
defined for $\alpha \le 1$. The scaling is defined and revised by
several authors (see for instance Ref.\cite{prbCannas} and
references therein). With such considerations we can repeat the mean
field approximation and we hope that the critical temperature comes
to be exact. In addition, the most important problem on the
thermodynamic behavior related to the systems with long range
interactions, it is the strong dependence on the size of systems. It
is thought that no standard thermodynamic equilibrium is reached;
then, it is crucial to make a suitable choice of the Hamiltonian.
However, we have presented a possible way to solve such problem. A
nice thermodynamic behavior is obtained from the Hamiltonian
suggested in Eq.(\ref{Hs}) to Eq.(\ref{Hs4}). Thermodynamic
quantities are not dependent on the size of the system for an
appropriated choice of the Hamiltonian.

Finally, in the study and characterization of the phase transition
and critical phenomena for systems with arbitrary range
interactions, advances and suggestions are always very important to
find appropriated models for calculating typical physical
quantities.

 It is a pleasure to acknowledge partial financial support by
 grants 1051075 and 1020269 from  FONDECYT and Milenio ICM
 P02-054F. In addition, we would like to thank to D. Barrios for his
 help in the initial implementation of Monte Carlo procedure.


\begin{thebibliography}{88}
\bibitem{PGaNat416} P. Gambardella, A. Dallmeyer, K. Maiti,
M.C. Malagoli, W. Eberhardt, K. Kern, and C. Carbone, Nature {\bf
416}, 301 (2002).
\bibitem{prlDorantes} J. Dorantes-Dávila and G. M.
Pastor, Phys. Rev. Lett. {\bf 81}, 208 (1998).
\bibitem{prbShen} J.Shen, R. Skomski, M. Klua, H.Jenniches, S.
Sundar Manoharan, and J.Kirschner, Phys. Rev. {\bf B 56}, 2340
(1997).
\bibitem{prlMermin} N.D.Mermin and H.Wagner, Phys. Rev. Lett.
{\bf 17}, 1133 (1966).
\bibitem{prbCannas} S. Cannas, F. Tamarit, Phys. Rev. {\bf B 54},
12661 (1996).
\bibitem{prb58} I. F. Herbut, Phys. Rev. {\bf B 58} 971 (1998).
\bibitem{PGaprb61} P. Gambardella, M. Blanc, H. Brune, K.
Kuhnke and K. Kern, Phys. Rev. {\bf B 61}, 2254 (2000).
\bibitem{Hove} L. van Hove, Physica {\bf 16}, 137 (1950).
\bibitem{Reichl} L. Riechl, {\em `` A Modern Course in Statistical"}
Mechanics, 2nd. Ed., John Wiley \& Sons, INC (1998).
\bibitem{SCuExact} S. Curilef, Physica {\bf A 344}, 456 (2004).
\bibitem{pre70} H. Chamati, D. Dantchev, Phys. Rev. {\bf E 70}, 066106
(2004).
\bibitem{mplb17} H. Chamati, N. S. Tonchev, Mod. Phys. Lett.
{\bf 17}, 1187 (2003); J. Phys. A: Math. Gen 33, L187 (2000).
\bibitem{pra45} R. Minieri, Phys. Rev. {\bf A 45}
3580 (1992).
\bibitem{prl76} E. Luijten and H. W. J. Bl\"ote, Phys.
Rev. Lett. {\bf 76}, 1557 (1996).
\bibitem{prl86} E. Luijten and H. Me$\beta$ingfeld, Phys. Rev. Lett. {\bf
86}, 5305 (2001).
\bibitem{prb56} E. Luijten and H. W. J. Bl\"ote, Phys. Rev. {\bf B
56}, 8945 (1997).
\bibitem{prb34} J. O. Vigfusson, Phys. Rev. {\bf B 34} 3466 (1986).
\bibitem{JPC4} P. W. Anderson and J. Yuval, J. Phys. {\bf C 4}, 607 (1971)
\bibitem{JPC3} J. F. Nagle and J. C. Bonner, J. Phys. {\bf C 3}, 352 (1970)
\bibitem{cssXII} E. Luijten, Computer Simulation Studies in
Condensed-Matter Physics XII, 86 (2000).
\bibitem{SCaIJMPC} S. Cannas, C. Lapilli and D. Stariolo, Int. J. Mod. Phys. {\bf C 15} 115 (2004).
\bibitem{note} For a discussion of the behavior of
this kind of expressions see for instance Ref.~\cite{prbCannas} and
references therein. We remark that, it has been originally proposed
as a nonextensive scaling. An alternative model is studied
here.
\bibitem{Lekner} J.Lekner, Physica {\bf A 157}, 826 (1989); Physica {\bf A
176} 485 (1991).
\bibitem{Jensen} N.Groenbech-Jensen, Int. J. Mod. Phys. {\bf C 6} 873 (1996).
\bibitem{efficient} H. Fangohr, A. Price, S. Cox, P. de Groot, G. Daniell, K. Thomas, J.
Comp. Phys. {\bf 162}, 372 (2004).
\bibitem{SCuIJMPC} S. Curilef, Int. J. Mod. Phys. {\bf C 11} 629 (2000).
\bibitem{Ruelle} D. Ruelle, {\em ``Statistical Mechanics"}, Imperial College Press and World Scientific (1999).
\bibitem{Binder} K. Binder and D. W. Heermann, {\em ``Monte Carlo
Simulation in Statistical Physics An Introduction"},  Fourth Edition
Springer (2002).
\bibitem{prb32} M. N Barber, R. B. Pearson, J. L. Richardson, D.
Touissant, Phys. Rev. {\bf B 32}, 1720 (1985).
\bibitem{tableIII} The reader can be addressed to the Table III of the
Ref.\cite{prb56} and to see previous results for $K_c$.
\end{thebibliography}
\end{document}